%% file: manuscript.tex
\documentclass[]{achemso}
\setkeys{acs}{articletitle = true}
\setkeys{acs}{etalmode = truncate}
\setkeys{acs}{maxauthors = 0}
\SectionNumbersOn

\usepackage[version=3]{mhchem}

\usepackage{siunitx}
\usepackage{multirow}
\usepackage{comment}
\usepackage[hidelinks]{hyperref}

\newcommand*\Eha{E_\mathrm{h}}
\newcommand*\vR{\mathbf{R}}
\newcommand*\vr{\mathbf{r}}
\newcommand*\diff{\mathrm{d}}
\newcommand*\MM{\mathrm{MM}}
\newcommand*\QM{\mathrm{QM}}
\newcommand*\QMMM{\mathrm{QM/MM}}

\DeclareSIUnit\angstrom{\text {Å}}
\DeclareSIUnit\bohr{\text {\ensuremath {a}}_{0}}

\newcommand{\EPFL}{Laboratory of Computational Chemistry and Biochemistry, École Polytechnique Fédérale de Lausanne, CH-1015 Lausanne, Switzerland}
\newcommand{\DTU}{DTU Chemistry, Technical University of Denmark (DTU), DK-2800 Kongens Lyngby, Denmark}

\author{Andrej Antalík}
\email{andrej.antalik@epfl.ch}
\author{Andrea Levy}
\author{Sophia K. Johnson}
\affiliation[EPFL]{\EPFL}

\author{Jógvan Magnus Haugaard Olsen}
\affiliation[DTU]{\DTU}

\author{Ursula Rothlisberger}
\email{ursula.rothlisberger@epfl.ch}
\affiliation[EPFL]{\EPFL}

\title{Making Puzzle Pieces Fit or Reshaping MiMiC for Multiscale Simulations with CP2K and More}

\begin{document}

\begin{abstract}
MiMiC is a framework for modeling large-scale chemical processes that require treatment at multiple resolutions.
It does not aim to implement single-handedly all methods required to treat individual subsystems, but instead, it relegates this task to specialized computational chemistry software while it serves as an intermediary between these external programs, and computes the interactions between the subsystems.
MiMiC minimizes issues typically associated with molecular dynamics performed with multiple programs, by adopting a multiple-program multiple-data paradigm combined with a loose-coupling model.
In this article, we present the addition of a new client program, CP2K, to the MiMiC ecosystem, which required a major refactoring of the entire framework and in the end allowed us to unlock its full flexibility.
By thorough timing analysis, we verify that the introduced changes do not affect
the performance of MiMiC or CP2K, and neither are they a source of significant computational overheads that would be detrimental to simulation efficiency.
Moreover, we demonstrate the benefits of the framework's modular design, by performing a QM/MM MD simulation combining CP2K with previously interfaced OpenMM, with no additional implementation effort required.
\end{abstract}

\section{Introduction}

% INTRO: MULTISCALE METHODS
Multiscale methods are a staple of computational modeling of large-scale systems undergoing complex chemical and physical processes involving different time and length scales.
One of the most prevalent approaches is the hybrid quantum mechanics/molecular mechanics (QM/MM)\cite{warshel1976theoretical, Singh_QMMM, Karplus_QMMM, Senn_QMMM, Rothlisberger_review, Estrin_review, Mennucci_review} model that combines a quantum mechanical (QM) description of a small subsystem with the rest of the system being described through classical molecular mechanics (MM) force fields.
It is a method of choice when studying processes that call for QM treatment taking place in a spatially limited region, with a need to account for environmental effects.
In such cases, applying QM methods to the entire system can be prohibitive and, more often than not, unnecessary.
Moreover, in the context of molecular dynamics (MD) based on ab initio methods, the use of QM/MM drastically reduces the computational cost compared to a full QM treatment of the entire system, thus allowing us to study processes on time scales that would otherwise be inaccessible due to system size.

% INTEGRATIVE FRAMEWORKS
Multiscale methods can be implemented in different ways, but typically a specialized software package is extended with the functionalities that it lacks, e.g., adding MM functionality to a quantum chemistry program.
A different strategy is to implement an integrative framework that merely handles communication and potentially computes subsystem interactions while the treatment of each subsystem is relegated to an external QM or MM program.
However, implementing such an approach can be rather intricate due to the need to maintain a suitable balance between flexibility and efficiency.
In the most straightforward approach, a flexible framework can be implemented in such a way that the external programs run without requiring any modifications while it handles data exchange by writing input files and reading output files.
However, this scheme suffers from large overheads due to slow data exchange via input/output operations and repeated startups and shutdowns of external programs, which are expensive, particularly in MD simulations run in highly parallel environments.

% MIMIC
MiMiC\cite{olsen2019mimic, antalik2024mimic} attempts to strike a balance between the two aspects of integrative frameworks by loosely coupling various external programs while maintaining an overall high simulation efficiency\cite{viacheslav2019extreme}.
The approach adopted for MiMiC is based on the multiple-program multiple-data (MPMD) paradigm with inter-program data transfer handled by a lightweight library that currently implements a mode of communication based on message-passing interface (MPI).
The entire simulation stack operates on a client--server model, in which subsystems of a studied system are assigned to individual external programs acting as clients, and the simulation workflow is controlled by a driver program acting as the server.
All programs are executed concurrently with communication managed by the server using a request-based approach.
In this setup, the server issues requests (send/receive data, run a calculation, etc.) that are received and executed by a client.
This allows the client programs to operate independently without any compromise in terms of resource allocation since MiMiC does not interfere with their underlying parallelization.
In terms of implementation, this approach also makes it easier for developers to couple their programs to MiMiC, as most of the interface can be written as stand-alone code.

% CP2K and WHY REFACTORING
The pilot implementation of MiMiC\cite{olsen2019mimic} featured CPMD\cite{cpmd_free} and GROMACS\cite{Abraham2015, van2005gromacs, berendsen1995gromacs} as QM and MM client programs, respectively, and we recently expanded the pool of supported software packages to OpenMM\cite{Levy2025} and Tinker-HP\cite{Sonata2024} as alternatives for the treatment of classical subsystems.
Owing to the MiMiC design, adding MM clients was (and continues to be) rather straightforward as writing an interface only requires on the client side.
However, the introduction of CP2K\cite{Kuhne2020} as a new QM client proved to be more challenging due to discrepancies between the design and the implementation as well as some remaining ties to CPMD, which has served simultaneously as a sole QM client and an MD driver until now.
Apart from writing an interface on the CP2K side, we refactored MiMiC and performed a careful disentanglement of MiMiC and CPMD in order to remove the remaining traces of tight coupling in MiMiC.

% PAPER OUTLINE
This article describes the modifications in the new refactored MiMiC framework and their impact on the performance of individual simulation components.
Section~\ref{sec:implementation} provides an overview of the framework and summarizes details about the modifications made to accommodate CP2K.
Section~\ref{sec:computational} introduces the systems used to study the performance of the refactored MiMiC framework with respect to key simulation parameters.
Section~\ref{sec:results} features simulation results that validate the stability of MD simulations performed with the refactored framework and presents an in-depth timing analysis that demonstrates its performance and impact (or rather lack thereof) on the performance of CP2K.

\section{Implementation Details}
\label{sec:implementation}

MiMiC is designed as a modular framework primarily for multiscale MD simulations of complex systems, where external programs are used to treat different subsystems.
The framework itself serves as an intermediary that facilitates data exchange and computes interactions that arise between the subsystems.
It is intended to be as flexible and efficient as possible, with minimal interference with the external programs and minimal effect on their performance.

\begin{figure}
\centering
\includegraphics[width=0.6\textwidth]{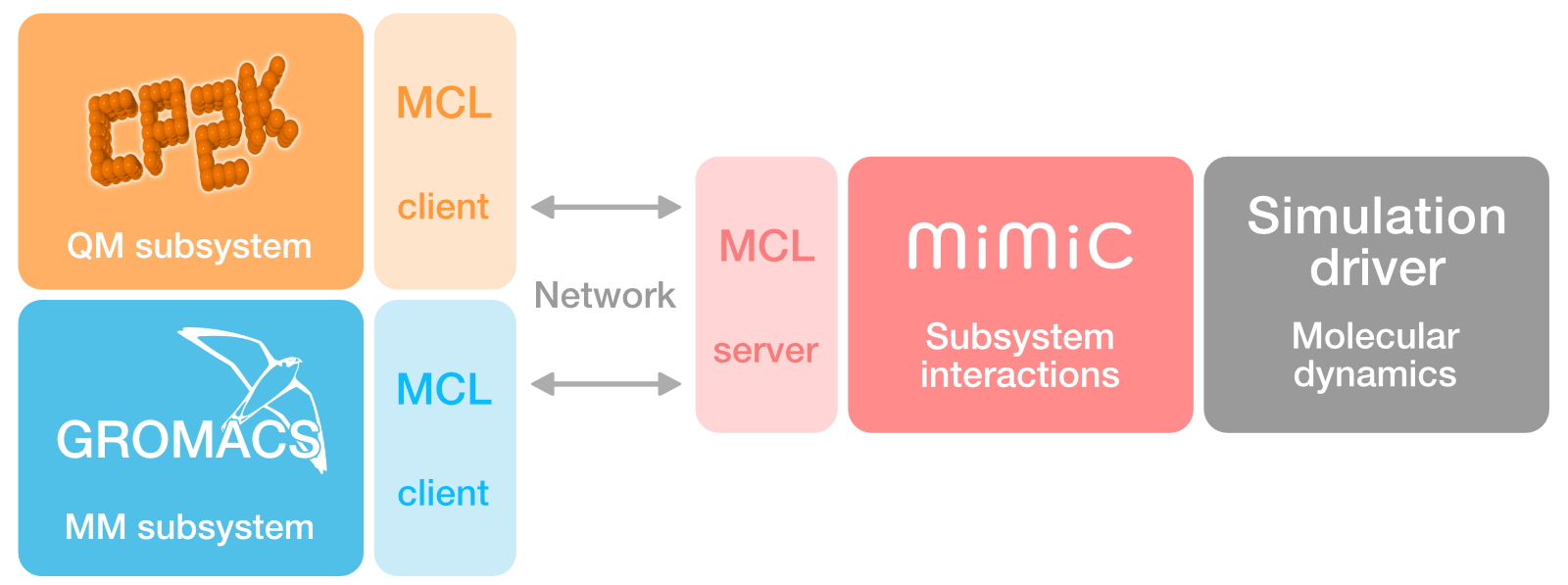}
\caption{Illustration of the strategy used by the MiMiC framework.}
\label{fig:general}
\end{figure}

For practical reasons, MiMiC is distributed in the form of two distinct libraries.
The main library with data structures and routines for managing subsystems and calculating their interactions is interfaced with the program that drives the MD simulation, i.e., integrates the equations of motion.
The external client programs only need to be linked against the lightweight communication library, MCL, which currently supports an MPI-based mode of communication.
Figure~\ref{fig:general} illustrates a typical setup with all the different components required for running a MiMiC-based QM/MM MD simulation involving CP2K and GROMACS.

A typical multiscale MD simulation with MiMiC starts with all programs launched and executed concurrently.
Right after the MPI world communicator is initialized in these programs, it is intercepted by an MCL routine that splits it and returns a local communicator for each program.
Next, invoking another MCL routine, a client--server relation between individual simulation components is established, with this relation being one of the core principles on which MiMiC operates.
In particular, the program interfaced with the main MiMiC library, i.e., the MD driver, acts as the server and issues requests to client programs, which are executed upon reception.
In the initial stage, the server collects all subsystem data from client programs and uses it to allocate data structures in the MD driver.
Then, in each MD step, the server distributes positions to programs handling individual subsystems and issues a request to start a force calculation.
If necessary, MiMiC computes subsystem interaction terms, such as energies, forces, and polarizing potential.
The step finishes with the server collecting energies and forces, which are then passed to the MD driver.
Finally, the velocities of all atoms constituting the system are updated, and their positions are propagated as classical point particles.

\subsection{Electrostatic embedding QM/MM in MiMiC}
\label{sec:es_emb}

Currently, MiMiC implements an additive QM/MM electrostatic-embedding scheme in which the point charges of the MM subsystem polarize the electron density of the QM subsystem.
In addition, van der Waals and certain bonded terms from the classical force field are included.
The electrostatic interactions are accounted for through explicit Coulomb interaction of classical point charges with QM nuclei and electron density.
Moreover, for increased efficiency, the interactions may be split into short- and long-range contributions based on distances between the MM atoms (or fragments) and the QM subsystem.
In the following paragraphs, we present the short- and long-range electrostatic interaction energies from which forces and polarizing potential are derived.
For more details on this approach, we refer to Ref. \citenum{olsen2019mimic}, which contains all relevant equations.

Short-range electrostatic interactions are calculated explicitly from the expression
\begin{equation}
\label{eq:v_es_sr}
    V^{\mathrm{es, sr}}_\QMMM = \sum_{i=1}^{N_\MM^{\mathrm{sr}}} q_{i}
    \left( \int \mathrm{d} \mathbf{r}\,   
    T_{\mathrm{mod}}^{(0)}(\vR_{\MM,i}, \mathbf{r}) \rho(\mathbf{r})
    + \sum_{j=1}^{N_\QM}  T_{\mathrm{mod}}^{(0)}(\vR_{\MM,i}, \vR_{\QM,j}) Z_j 
    \right) \ ,
\end{equation}
where $q_i$ is the effective point charge of the $i$th MM atom, $\rho$ is the electron density, and $Z_j$ is the charge of the $j$th QM nucleus.
Coulomb interactions are accounted for by the zeroth-order modified interaction tensor $T_{\mathrm{mod}}^{(0)}(\vR_a, \vR_b)$, which tends to $r^{-1}_{\mathrm{cov},\, a}$, i.e., the inverse covalent radius of atom $a$, as $|\vR_b - \vR_a| \rightarrow 0$, to prevent overpolarization of the QM subsystem and potential electron spill-out\cite{laio2002hamiltonian}.
If this approach is applied to the entire system, i.e., all MM atoms are treated as short-range, the electrostatic subsystem interactions are accounted for exactly.
Yet, this becomes expensive rather quickly with system size due to the repeated integration over the electron density grid for each MM atom.

To remedy this steep increase in computational cost, MiMiC sorts MM atoms into short- and long-range groups with respect to a user-specified cutoff distance.
While the former interacts with the QM subsystem explicitly according to Eq.~\ref{eq:v_es_sr}, the latter does so through a multipole expansion of the QM potential
\begin{equation}
\label{eq:v_es_lr}
    V^{\mathrm{es, lr}}_\QMMM = \sum_{i=1}^{N_\MM^{\mathrm{lr}}}\sum_{\left|\alpha\right|=0}^{A_\QM}
    \frac{(-1)^{\left|\alpha\right|}}{\alpha!}
    q_i  T^{[\alpha]}(\vR_{\MM,i}, \bar{\vR}_\QM) M_\QM^{[\alpha]} \ ,
\end{equation}
where $T^{[\alpha]}$ is the (non-modified) interaction tensor of $\alpha$th order (multi-index notation).
The remaining parameters, $\bar{\vR}_\QM$ and $A_\QM$, are the center and order of the multipole expansion, whose components are evaluated as
\begin{equation}
\label{eq:multipole_lr}
 M_\QM^{[\alpha]} = 
 \int \diff\vr \, \rho(\vr){\left( \vr-\bar{\vR}_\QM \right)}^\alpha
 + \sum_{j=1}^{N_\QM}Z_j {\left( \vR_{\QM,j}-\bar{\vR}_\QM \right)}^\alpha \ .
\end{equation}
This means that the long-range MM atoms interact only with the multipoles of the QM charge density, which only need to be evaluated once per MD step and only require a single integration of the electronic density per multipole component.
Thus, by carefully selecting the parameters, i.e., cutoff distance, multipole expansion order, sorting method, and sorting rate, users can tune their simulations to perform as efficiently as possible with virtually the same accuracy as a full short-range description.

\subsection{Modifications in MiMiC}

To accommodate other external QM programs than CPMD, MiMiC had to undergo several modifications.
The reason for this was that the pilot implementation\cite{olsen2019mimic, viacheslav2019extreme}, which included interfaces to CPMD and GROMACS, had a rather high degree of entanglement between MiMiC and CPMD, partly due to CPMD being used as an integrated QM program and MD driver.
In fact, many of the data structures in MiMiC, together with its approach to parallelization, directly followed those of CPMD\cite{viacheslav2019extreme}.
This helped to minimize the data exchange between individual software components and reduced the number of expensive parallel operations, such as all-to-all reductions, when handling distributed data.

The original implementation was used to run simulations with GROMACS as the MM program while CPMD served as both MD driver and QM program, with MiMiC often sharing the same data structures with CPMD.
In the new implementation, CPMD can still be used as an MD simulation driver and also, separately, as a QM client program.
Now having a QM subsystem treated by a program running independently of the MD driver, MiMiC has to allocate memory for data storage of the relevant quantities.
Therefore, the first step was to introduce the initial collection of QM subsystem information, with the subsequent transfer of positions and forces at every step of an MD simulation.
By eliminating MiMiC--CPMD interdependencies, MiMiC now stores its own copy of relevant data and becomes the sole simulation component that maintains complete information about the full system.

In addition, we also wanted to enhance flexibility of the framework for future developments, so we decided to reinvent the way MiMiC handles QM/MM interactions in the electrostatic-embedding scheme and, in particular, stores grid quantities, i.e., external polarizing potential and electron density.
We approached this with two objectives in mind.
Namely, to be more flexible by enabling the use of non-uniform grids, e.g., adaptive grids for programs using a finite-element basis, while, at the same time, to remain as computationally efficient as in the original implementation.
To address the first objective, we serialized the storage of these quantities by keeping grid point coordinates with their corresponding values.
However, the increased flexibility came with a price, since replacing the shared data allocations of MiMiC and a QM client program substantially increased the amount of data transferred in each simulation step.
Such data is often distributed over many processes, meaning that it needs to be collected at/scattered from the respective root processes before being transmitted between programs.
These operations can introduce non-negligible overhead, and for this reason, we assess them through benchmark simulations.
On the other hand, the new approach gives us more freedom in terms of parallelization as we are no longer bound by the grid dimensions, and we can distribute the grid into as small chunks as necessary.

\subsection{CP2K Interface}

The changes described in the previous subsection were motivated by our intent to introduce CP2K into the MiMiC ecosystem.
This also required implementing an interface in CP2K, but thanks to CP2K's modular design, this interface is well-separated from the rest of the code.
Practically, the only part of the code that modifies a core CP2K routine is the splitting of the MPI world communicator, which has to be intercepted right after its initialization \verb|MPI_Init| by invoking \verb|MCL_Init|.
The rest of the interface that takes care of communication via MCL is written in independent modules \verb|mimic_loop| and \verb|mimic_communicator|.

After launching an MD simulation, invoking the newly introduced run type in CP2K called 'MIMIC' steers the program flow toward the \emph{MiMiC loop}, a central construct for a client's operation.
It is a conditional loop in which the client program receives and executes a request from the server in each iteration (for more details, see Ref.~\citenum{antalik2024mimic}).
The actual communication is done through the routines implemented in the communicator module that fetch and transfer relevant data required for running a simulation to the server, e.g., all system information during the initialization phase.
A typical step of an MD simulation then consists mainly of receiving the positions and the external polarizing potential from the server, calling the routine that solves the QM equations and evaluates forces, and transferring the final electron density, energies, and forces back to the server.
Once the program receives a request to finalize the simulation, the program exits the loop and terminates its execution.

\section{Computational Details}
\label{sec:computational}

\begin{figure}[ht]
    \centering
    \includegraphics[width=0.5\linewidth]{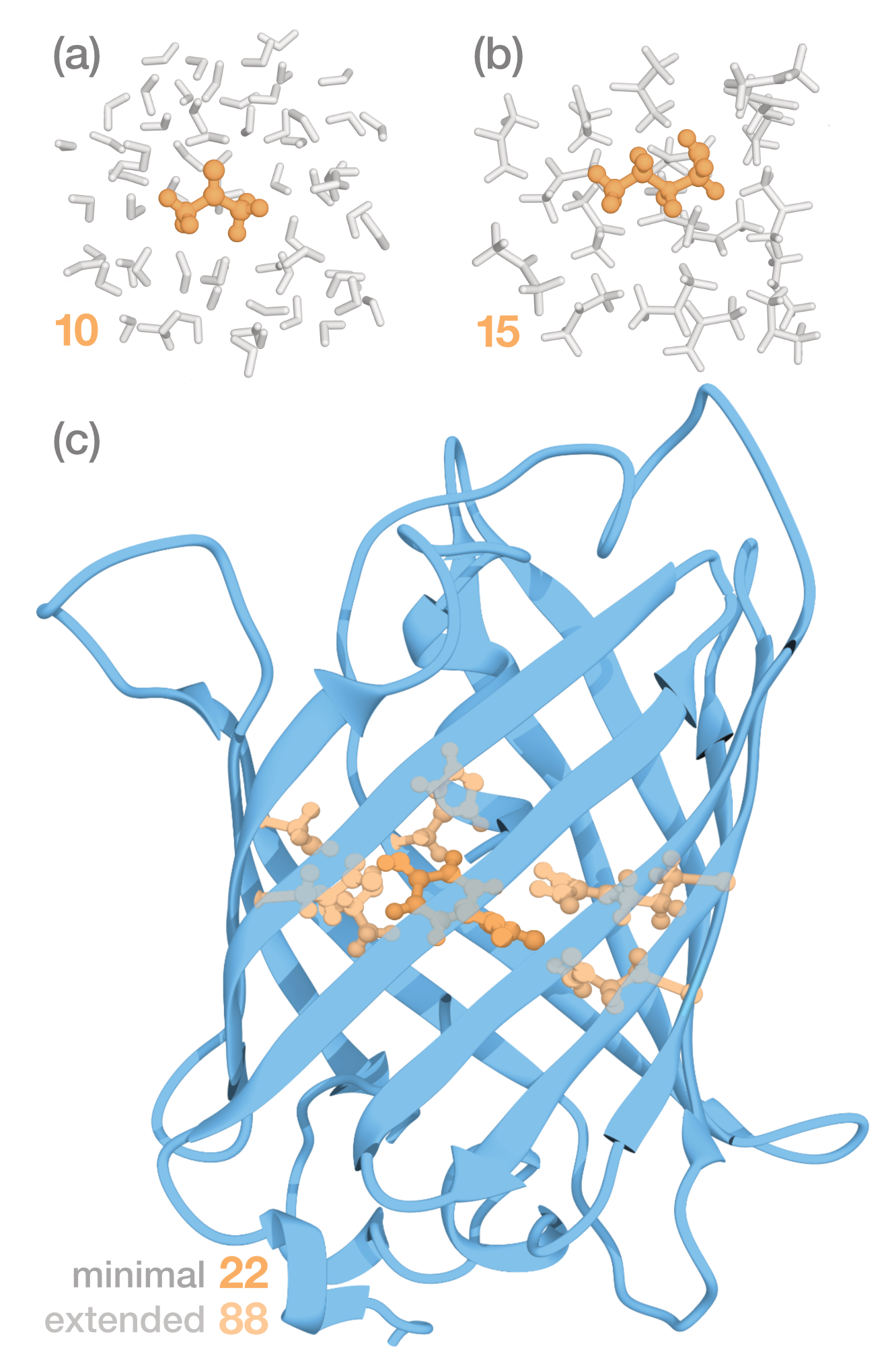}
    \caption{Systems used in this work: (a) acetone solvated in water (Ace(aq)), (b) butanol solvated in acetone (BuOH(ace)), (c) green fluorescent protein solvated in water (GFP(aq)).
    Orange ball-and-stick models represent the QM subsystems, blue ribbons the protein, and gray licorice the solvent (in GFP(aq) omitted for clarity).
    The orange numbers represent the number of atoms in the QM subsystem.}
    \label{fig:systems}
\end{figure}

% VALIDATION: SYSTEM DESCRIPTION
We performed validation simulations with three different systems of increasing size, namely, an acetone molecule solvated in \num{978} water molecules (Ace(aq)), green fluorescent protein solvated in \num{8219} water molecules (GFP(aq)), and butanol solvated in \num{13948} acetone molecules (BuOH(ace)) (see Figure~\ref{fig:systems}).
We chose these systems to represent typical situations that might occur during simulations.
Ace(aq) as the simplest one consists of a single acetone molecule of \num{10} atoms, which constitutes the QM subsystem, solvated in rigid water molecules modeled at the classical level, thus totaling in \num{2944} atoms.
The main difference in the case of BuOH(ace), apart from its size, is the fact that we use acetone as a solvent, whose molecules remain unconstrained during the simulation.
The QM subsystem is composed of a single butanol molecule of \num{15} atoms, with the remainder being treated at the MM level with \num{139495} atoms in total.
GFP(aq) is the most biologically relevant system used in validation, whose QM subsystem contains its neutral 4-hydroxybenzylidene-2,3-dimethylimidazolinone chromophore of \num{22} atoms, which include 2 boundary atoms to account for the bonds crossing the QM--MM boundary.
Its MM subsystem is then composed of the rest of the protein, solvent, and counter ions resulting in a total of \num{28264} atoms.

% VALIDATION: SYSTEM PREPARATION
For all systems, we took the initial pre-equilibrated (at QM/MM level) configurations from repositories.
In particular, we selected the two solute--solvent systems from a previous work on MiMiC\cite{olsen2019mimic}, while the GFP(aq) system was chosen out of the BioExcel Benchmark Suite\cite{github_bioexcel_benchmarks} (PDB ID: 1GFL\cite{yang1996molecular, pdb_1gfl}).
To ensure equal treatment, we further equilibrated the systems by performing \num{45000} step of Born--Oppenheimer QM/MM MD with a time step of \num{20.0}~$\hbar/\Eha$ ($\sim$\SI{0.48}{\femto\second}) in the NVT ensemble using a Nosé-Hoover thermostat set to \qty{298.15}{\kelvin}.
This was followed by $\sim$\SI{9.6}{\pico\second} NVE runs, with the first part of the simulation being considered a part of equilibration and only the last \SI{5}{\pico\second} used for analysis.

% MM PARAMETERS
We carried out the simulations describing the classical part of the studied systems with a modified version of GROMACS v2021.6\cite{Abraham2015, van2005gromacs, berendsen1995gromacs}. 
In the case of Ace(aq) and BuOH(ace) the MM subsystems were modeled using the OPLS/AA force field\cite{Jorgensen2005} in conjunction with the TIP3P water model\cite{Jorgensen1983}.
The same water model was employed in the GFP simulations while the protein and counter ions were described by the AMBER ff03 force field\cite{Duan2003}.
The covalent bonds of water molecules in Ace(aq) and GFP(aq) systems were constrained using the RATTLE algorithm \cite{Andersen1983, Weinbach2005}.
The simulation box dimensions were unaltered compared to the inputs we acquired from the repositories.
Further details can be found in Supporting Information (SI) Table~\ref{tab:gromacs_params}.

% QM PARAMETERS
We performed simulations with either CPMD v4.3\cite{cpmd_free} modified to be compatible with MiMiC v0.2.0 (\emph{previous} version) or with CP2K v2023.1 (our development branch) with a development version of MiMiC (\emph{refactored} version).
Validation and timing simulations, whose results are presented in Sections \ref{sec:valid} and \ref{sec:timings}, were performed using the BLYP functional\cite{Becke1988,Lee1988}, but several different functionals/methods (described below) were used in addition to BLYP for CP2K performance assessment. 
%CPMD
For simulations carried out with CPMD, we used Troullier--Martins norm-conserving pseudopotentials\cite{Troullier1991}, and the Martyna--Tuckerman Poisson's equation solver\cite{Martyna1999} to model the QM subsystem with isolated boundary conditions.
For Ace(aq) and BuOH(ace), we used the QM box sizes and plane wave (PW) basis-set cutoffs from previous work\cite{olsen2019mimic}, while in the case of GFP we based our settings on benchmark calculations, and these were used in all calculations involving CPMD (see Table~\ref{tab:cpmd_mimic_params} in SI).
%CP2K
Simulations in CP2K\cite{Kuhne2020} were performed with the Quickstep module for electronic-structure calculations\cite{VandeVondele2005quickstep} using the mixed Gaussian and PW basis-set approach \cite{Lippert1997} using five multigrids.
Here we employed Goedecker--Teter--Hutter pseudopotentials \cite{Goedecker1996,Krack2005} with double- and triple-$\zeta$ Gaussian basis sets that were optimized for molecular systems by \citeauthor{VandeVondele2007}\cite{VandeVondele2007}.
Periodic images were decoupled using the density-derived atomic point charges\cite{Blochl1995}.
To saturate the bonds in GFP that cross the QM--MM boundary, we replaced $\mathrm{C}_\alpha$ atoms with monovalent pseudopotentials by \citeauthor{komin_capping_2009}\cite{komin_capping_2009} for simulations with both CPMD and CP2K.
For validation, unlike with CPMD, we used two different sets of cutoffs, basis sets, and box sizes, labeled \emph{loose} and \emph{tight} parameters, both described in SI, Table~\ref{tab:cp2k_params}.
For equilibration in the NVT ensemble, we used the loose settings based on the parameters set automatically by the existing QM/MM interface with CP2K in GROMACS, developed by the BioExcel Centre \cite{bioexcel_qmmm}.
For NVE validation runs, we used the tight set of parameters that we determined by carefully benchmarking PW cutoffs and simulation box size.
% MiMiC PARAMETERS
Long-range interaction parameters, i.e., short-range cutoff and multipole expansion order, in MiMiC were either taken from the previous article for solute--solvent systems\cite{olsen2019mimic} or benchmarked for GFP(aq) (see SI).
The short- and long-range atom groups were updated every 50 steps, with atom-wise fragment sorting for GFP(aq) and centroid-based sorting for the other systems.

% TIMINGS: BRIEF SYSTEM DESCRIPTION + ADDITIONAL PREP
The timing analysis presented in Section~\ref{sec:timings} was performed for several simulations on systems derived from the Ace(aq) system.
In addition to the system setup already described above, which is referred to as \emph{Small MM}, we set up another, \emph{Large MM}, in which the acetone molecule was solvated in a larger box with 
\num{32768} water molecules leading to a full system comprising \num{98314} atoms in total.
Just like the smaller system, it was equilibrated with the same QM region for \num{45000} steps in an NVT ensemble simulation using the loose set of parameters for CP2K.
Taking these equilibrated structures, we defined \emph{Small QM} and \emph{Large QM} systems (described in the results section) and, for each combination of small and large QM and MM subsystems, performed a short equilibration NVE run of \num{2500} steps with the respective set of parameters from Table~\ref{tab:cp2k_params}.
The final configurations were then used to run simulations to generate data for the timing analysis.

% PERFORMANCE: SYSTEM DESCRIPTION, PARAMETERS
For the client--program performance assessment presented in Section~\ref{sec:performance}, we once again used GFP(aq), but with a larger QM subsystem comprising \num{88} atoms, including \num{4} boundary atoms.
Apart from the previously used BLYP functional, here we also employed B3LYP\cite{Stephens1994, Becke1988, Lee1988, Vosko1980} as a representative case of a hybrid functional, CAM-B3LYP\cite{Yanai2004,Shao2019} as a case of a range-separated hybrid functional, and second-order Møller--Plesset perturbation theory\cite{DelBen2012} (MP2) as an example of a post-Hartree--Fock method.
To accelerate the calculation of exact exchange, we employed the auxiliary density matrix method\cite{Guidon2010} (ADMM) for B3LYP and CAM-B3LYP, and the resolution-of-identity (RI) approximation\cite{DelBen2013} for MP2.
To set up these simulations, we started from the structure that was equilibrated in simulations with the small QM subsystem (22 atoms) and performed a $\sim$\SI{4.8}{\femto\second} NVT run with the BLYP functional together with the loose settings in CP2K.
This was followed by a $\sim$\SI{3.8}{\femto\second} NVE run and the final configuration was used for a short production run of no more than 20 MD steps (further detailed in Sec.~\ref{sec:results}).

% MODULARITY
Finally, we also performed additional simulations on GFP(aq) with the small QM subsystem, where we replaced the MM program with OpenMM v8.0.0\cite{Eastman2023openmm,openmm:8.0.0} exploiting the recently published interface with MiMiC\cite{Levy2025}.
Starting from the final conformation of the system used during the validation, we performed a QM/MM MD simulation for $\sim$\SI{7.2}{\pico\second} in the NVE ensemble and used the last \SI{5}{\pico\second} for analysis.

% Hardware
We performed all simulations on the Piz Daint supercomputer at the Swiss National Computing Centre (CSCS).
Each node features an Intel Xeon E5-2690 12-core CPU with an NVIDIA Tesla P100 GPU accelerator.
For all simulations presented here, we refrained from using OpenMP threading in MiMiC and made use solely of the MPI parallelism, meaning that a single MPI process ran using a single CPU core.
We made use of GPUs in QM subsystem calculations with CP2K, as supported in the v2023.1.

\section{Results and Discussion}
\label{sec:results}

\subsection{Validation}
\label{sec:valid}

To validate the refactored MiMiC framework and the interface in CP2K, we first studied the total energy fluctuations in QM/MM NVE MD simulations to verify that it is indeed a conserved quantity.
We performed these simulations with three different systems of increasing size, which also represent typical situations encountered in QM/MM MD simulations.
Two solute--solvent systems, Ace(aq) and BuOH(ace), differ mainly in size, but the latter also involves a solvent whose molecules remain unconstrained throughout the simulation.
The last system, GFP(aq), is probably the most interesting for the deployment of MiMiC to study processes of potential biological relevance, as it involves covalent bonds across the QM--MM boundary, which is a rather common feature in these kinds of systems.

\begin{figure}
\centering
\includegraphics[width=0.5\textwidth]{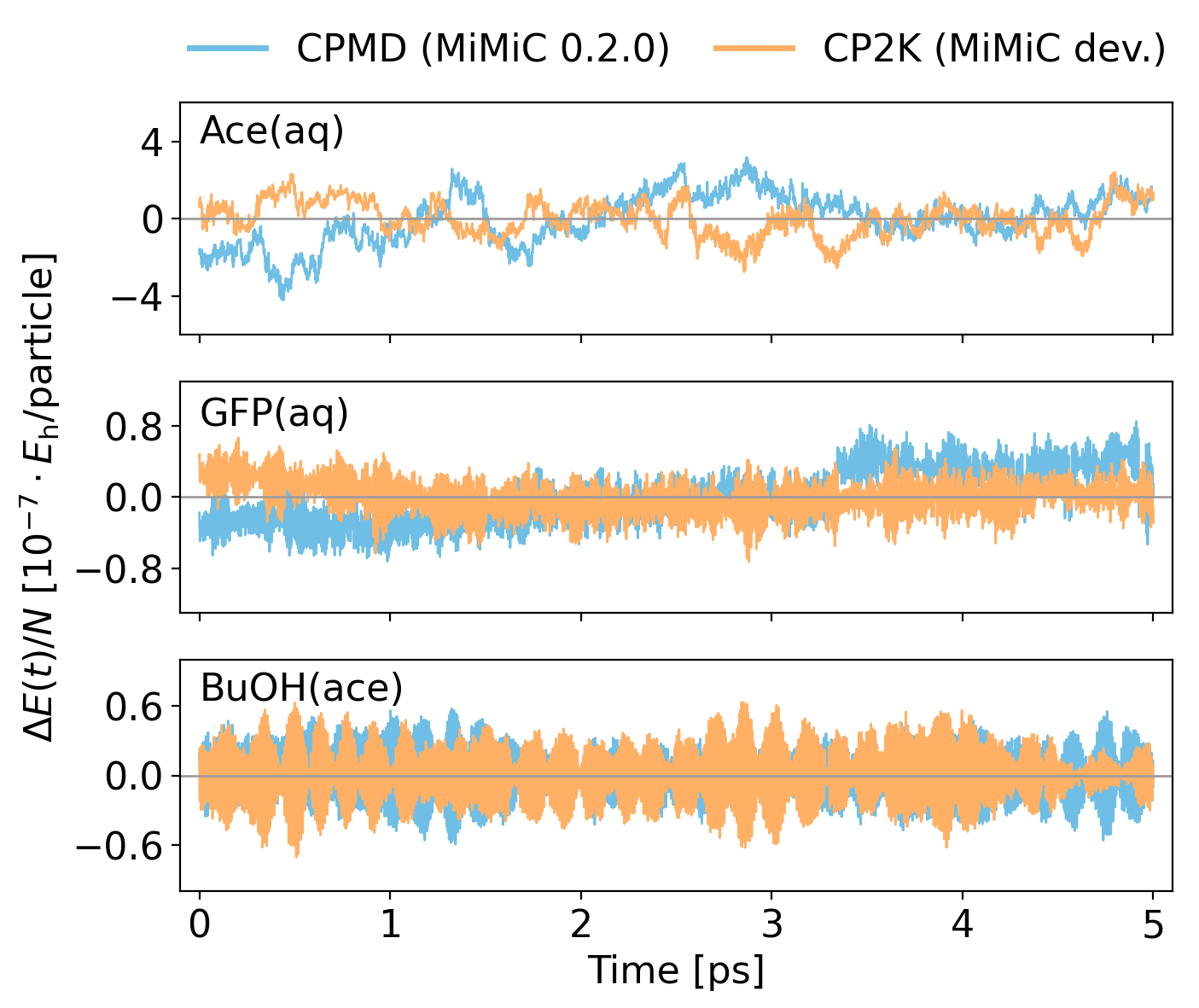}
\caption{Fluctuations of the total energy per particle in QM/MM MD simulations performed with MiMiC for Ace(aq), GFP(aq), and BuOH(ace).
The change in energy per particle is given relative to the average energy ($\Delta E(t) = E(t) - \langle E \rangle$).
The orange lines correspond to the simulations performed with the refactored development version of MiMiC using CP2K to treat the QM subsystem, while the blue lines come from simulations performed with the previous version of MiMiC with the QM subsystem handled by CPMD.
All simulations used the long-range coupling scheme for the electrostatic QM/MM interactions (see Sec.~\ref{sec:es_emb}).
}
\label{fig:validation}
\end{figure}

The results from these simulations are shown in Figure~\ref{fig:validation}, which features fluctuations in total energy per particle shown with respect to the average energy.
They were performed with the refactored version of MiMiC using CP2K to treat the QM subsystem and, for comparison, with the previous version of MiMiC using CPMD.
We employed the electrostatic embedding scheme described in Section~\ref{sec:es_emb} including the long-range coupling.
Neither of the studied systems exhibits any observable drift, as all of them appear to fluctuate around the mean value.

Since the refactoring also affects the earlier interface with CPMD, in addition to these results, we performed several simulations with the refactored version of MiMiC using CPMD to treat the QM subsystem.
They can be found in the SI, together with results from the NVE simulations using CP2K with the loose settings, originally used only in NVT.

\subsection{Data transfer assessment}
\label{sec:timings}

In its pilot implementation\cite{olsen2019mimic,viacheslav2019extreme}, the data transfer between MiMiC and clients only involved transferring positions and energies with the MM client.
For the transfer of other quantities between MiMiC and CPMD, the two shared some data structures, which reduced the communication overhead effectively preventing it from becoming a bottleneck.
However, this also meant that MiMiC was not entirely decoupled from CPMD, thus replacing it with a different QM program was not straightforward and required ad-hoc modifications in MiMiC.
The refactoring changed this but also introduced additional communication of quantities between MiMiC and QM clients, like electron densities or external potentials, which in PW-based programs are expressed on grids of significant sizes.
Therefore, we investigate the impact of this additional data transfer on the overall performance of MiMiC-based MD simulations.

For this purpose, we modeled four situations by setting up four systems based on Ace(aq) representing combinations of small and large MM subsystems with either a small or a large QM subsystem.
Here, we define the sizes from the PW-based perspective, i.e., the MM subsystem size is determined by the number of classical atoms, while for the QM subsystem, the relevant quantity is the size of the real-space grid used to store density and external polarizing potential.
Regarding the MM subsystem size, the small one contains as many atoms as the one used during validation, while the large one has over 30 times as many water molecules. 
For the QM subsystems, we defined a small QM subsystem with a grid of $93^3$ grid points and an artificially large one with $384^3$ grid points by adjusting cutoffs and QM box sizes.
This approach is reasonable since in PW-based calculations (QM or QM/MM), it is the size of the grid, not the number of atoms, that is the leading factor that impacts scaling.

\begin{figure}
    \centering
    \includegraphics[width=\linewidth]{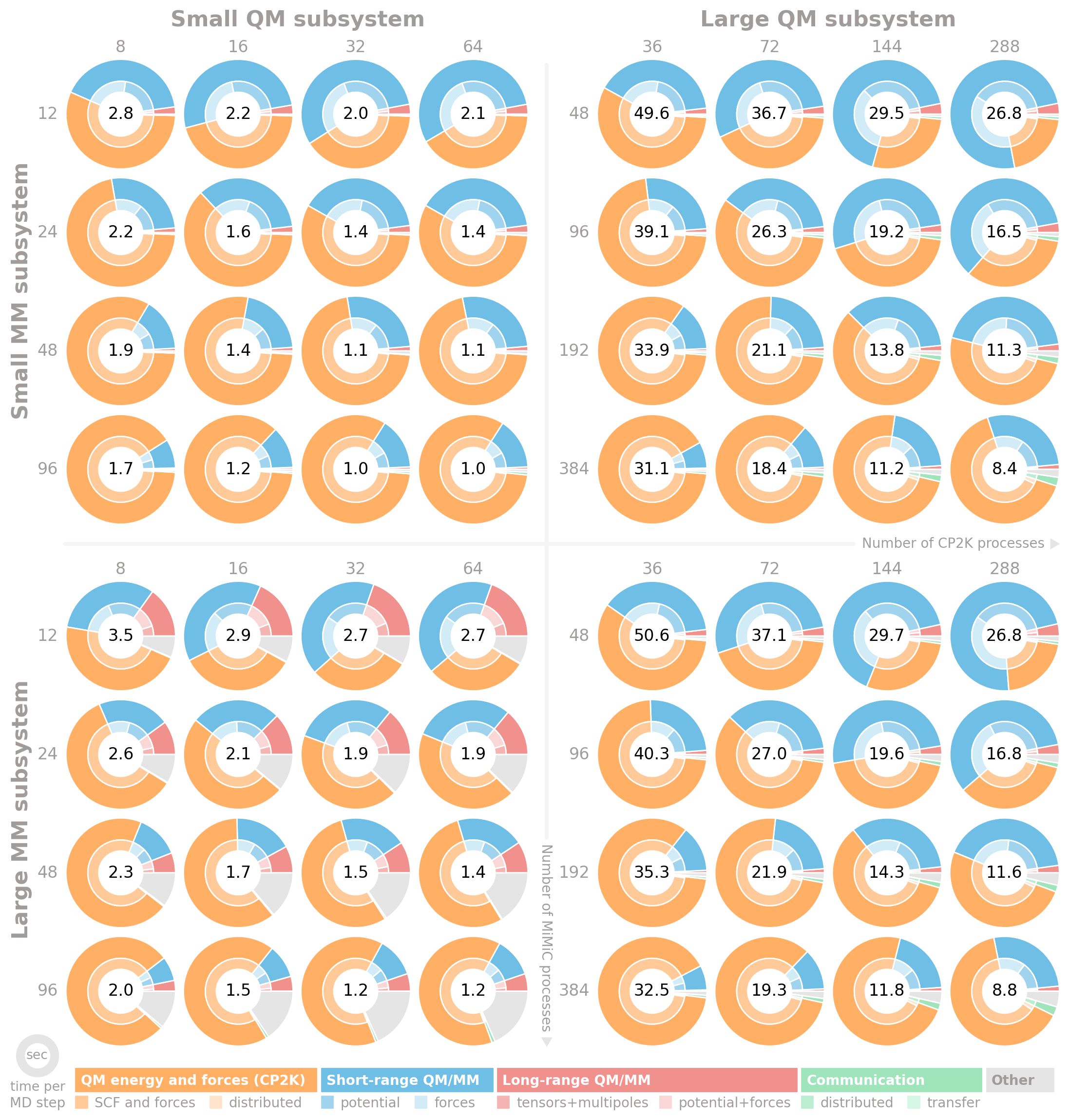}
    \caption{Timing analysis of MiMiC-based QM/MM MD simulations of systems with an acetone molecule in the QM subsystem solvated in classical MM water molecules.
    The plot features different combinations of small and large subsystems, which, in the case of MM, relates to the number of solvent molecules and, in the case of QM, the number of grid points in the real-space grid.
    The donut charts are ordered on the grid with respect to the increasing number of MPI processes for CP2K (horizontal) and MiMiC (vertical).
    The \emph{distributed} items refer to the gathering to and scattering from the rank-0 processes of distributed data structures before and after its transmission between a client and the server.
    The number in the circle is the total time per MD step in seconds.
    Note that on top of CPUs, CP2K used an additional GPU per 8 MPI processes.}
    \label{fig:data_transfer}
\end{figure}

Figure \ref{fig:data_transfer} shows the timing analysis for increasingly distributed calculations in the form of donut charts, which represent the portion of time spent on different actions within an MD step, with the total time per MD step printed in the center.
Each quadrant features a set of these charts for a simulation performed with an increasing number of MPI processes for CP2K (horizontal) and MiMiC (vertical).
We calculated these timings as averages over 20 MD steps, except for the CP2K SCF solution and force evaluation, in which we omitted the first four steps where the wave function extrapolation is not yet fully engaged.

Starting with the smallest system, we observe a decrease in the time per time step with an increasing number of processes, as expected.
Due to the modest size of this system, simulation performance gets saturated rather quickly on CP2K's side, with virtually no improvement above 16 MPI processes.
Yet, MiMiC preserves its scaling almost perfectly with parallel efficiency over \num{97}\,\% with respect to the least distributed calculation, i.e., running with 12 MPI processes.
In the system with a large MM subsystem and small QM subsystem, a much larger number of classical atoms interact with the QM subsystem through long-range interactions.
Consequently, the relative times spent in different parts of the simulation change, with a more prominent portion now spent on calculating the long-range potential and forces (note that the long-range approximation parameters remain the same).
Also, the section `Other', which groups actions such as the integration of the equations of motion, now becomes more noticeable, with this increase being mainly attributed to the constraint solver in CPMD, which is rather inefficient for a large number of constraints.
Still, compared to the smaller system, the computational overhead per MD step is no more than 33\,\% even though the size of the system increased more than 30-fold.
For the large QM subsystem, CP2K does not get saturated as compared to the small one, at least not within the studied range of processes.
Another difference is the communication that now becomes noticeable in the donut charts.
It is dominated by the collection and distribution of distributed data such as the electron density, with the time for data transfer being negligible.
Interestingly, this time remains practically constant, regardless of the number of processes used for MiMiC or CP2K.
In any case, the communication overhead does not constitute more than 5\,\% of the overall wall time per MD step in all of the studied cases.
Finally, the large MM subsystem combined with the large QM subsystem exhibits only minimal computational overhead of no more than 6\,\% in terms of wall time, with respect to the simulations performed with the same QM subsystem but a small MM subsystem.

Considering that the refactoring may affect the parallel scaling, in addition to these tests, we assessed the parallel performance of MiMiC alone to ensure it is not degraded compared to the previous version\cite{viacheslav2019extreme}.
For this, we picked the system with large QM and MM subsystems, fixed the number of CP2K processes at \num{144}, and altered the number of MiMiC processes up to a total of \num{8192}.
We performed these simulations with four processes per node for most simulations, with the exception of one- and two-process runs.

\begin{figure}
\centering
\includegraphics[width=0.5\textwidth]{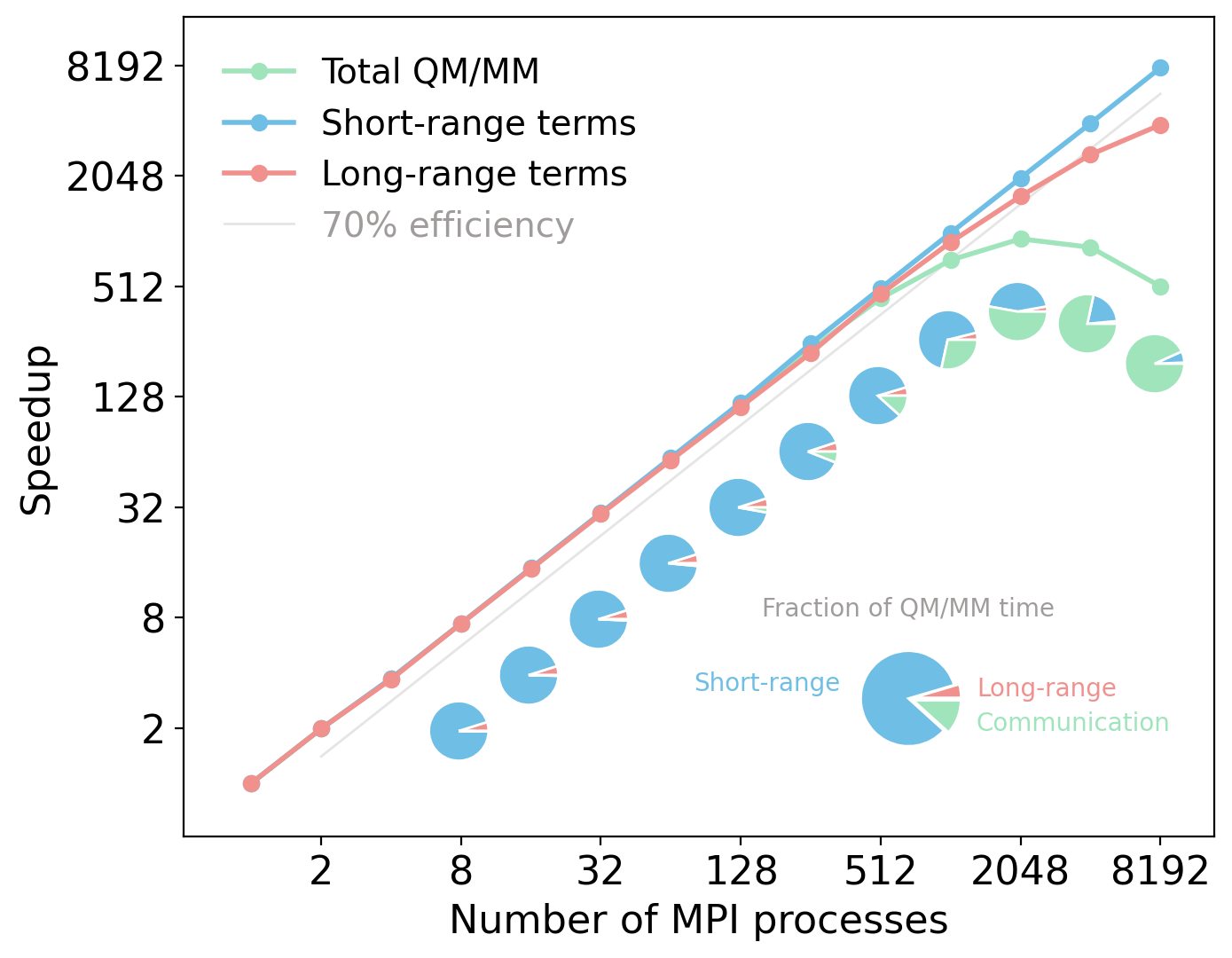}
\caption{Speedup plot demonstrating the parallel efficiency of MiMiC in the evaluation of QM/MM terms.
Calculations were performed for the Ace(aq) system whose QM subsystem was simulated in a cubic box with $384^3$ grid points and solvated in \num{32768} classically modeled water molecules.
Ideal scaling is achieved for the computation of short-range interactions.}
\label{fig:mimic_scaling}
\end{figure}

Figure~\ref{fig:mimic_scaling} shows speedups of MiMiC QM/MM routines, which are reported with respect to single-process performance.
Owing to its embarrassingly parallel nature, the computation of short-range interaction terms scales perfectly up to the maximum number of processes without any decrease in parallel efficiency.
The computation of long-range interactions also scales well but the efficiency drops to 65\,\% at \num{4096} MPI processes.
However, the overall performance of QM/MM hits the 70\,\% boundary much sooner at \num{1024} processes.
A detailed look at the timing decomposition shows that with \num{1024} processes the communication accounts for more than a quarter of the total time spent on QM/MM.
This fraction further grows with the number of processes and at \num{8192} the communication corresponds to almost the entire time spent on QM/MM.

While the communication is a bottleneck for the QM/MM part, it is insignificant for the total time per MD step in the cases investigated here.
For one, treating the QM subsystem is usually the most expensive part of a QM/MM simulation, and QM programs typically reach saturation at a point where wall times are still much longer than the time spent on QM/MM.
Therefore, increasing resources allocated to MiMiC yields only minuscule gains in wall times.
For example, gaining \SI{0.5}{\second} on the evaluation of QM/MM is not worth twice the resources when the QM program reaches saturation at \SI{5}{\second} per MD step.
In the near future, we also plan to offload the computation of short-range interactions to GPUs, which should further speed up the simulations.

\subsection{Performance and modularity}
\label{sec:performance}

In this section, we show that the performance of CP2K is practically unaffected by MiMiC, which is in line with our philosophy of minimal interference with the normal execution of external programs.
To demonstrate this, we used GFP(aq) with a larger QM subsystem and equilibrated it at the BLYP level.
Then, we took the final configuration and compared the scaling of BLYP, B3LYP, CAM-B3LYP, and MP2.
For each method, we ran two types of MD simulations: a QM/MM one using MiMiC, and another driven by CP2K with the QM subsystem only.
In both, we always tracked the same times, i.e., like in the previous section, just those needed to solve the QM problem and evaluate QM forces.
We evaluated the timings by averaging over \num{11} steps with the exception of the MP2 runs, where we averaged only over \num{6} steps.
We made this choice because a single-node MP2 run yielding \num{15} steps (first 4 discarded, see previous section) took longer than \num{24} hours which was the wall-time limit on the used supercomputer.

\begin{figure}[t]
    \centering
    \includegraphics[width=1.0\linewidth]{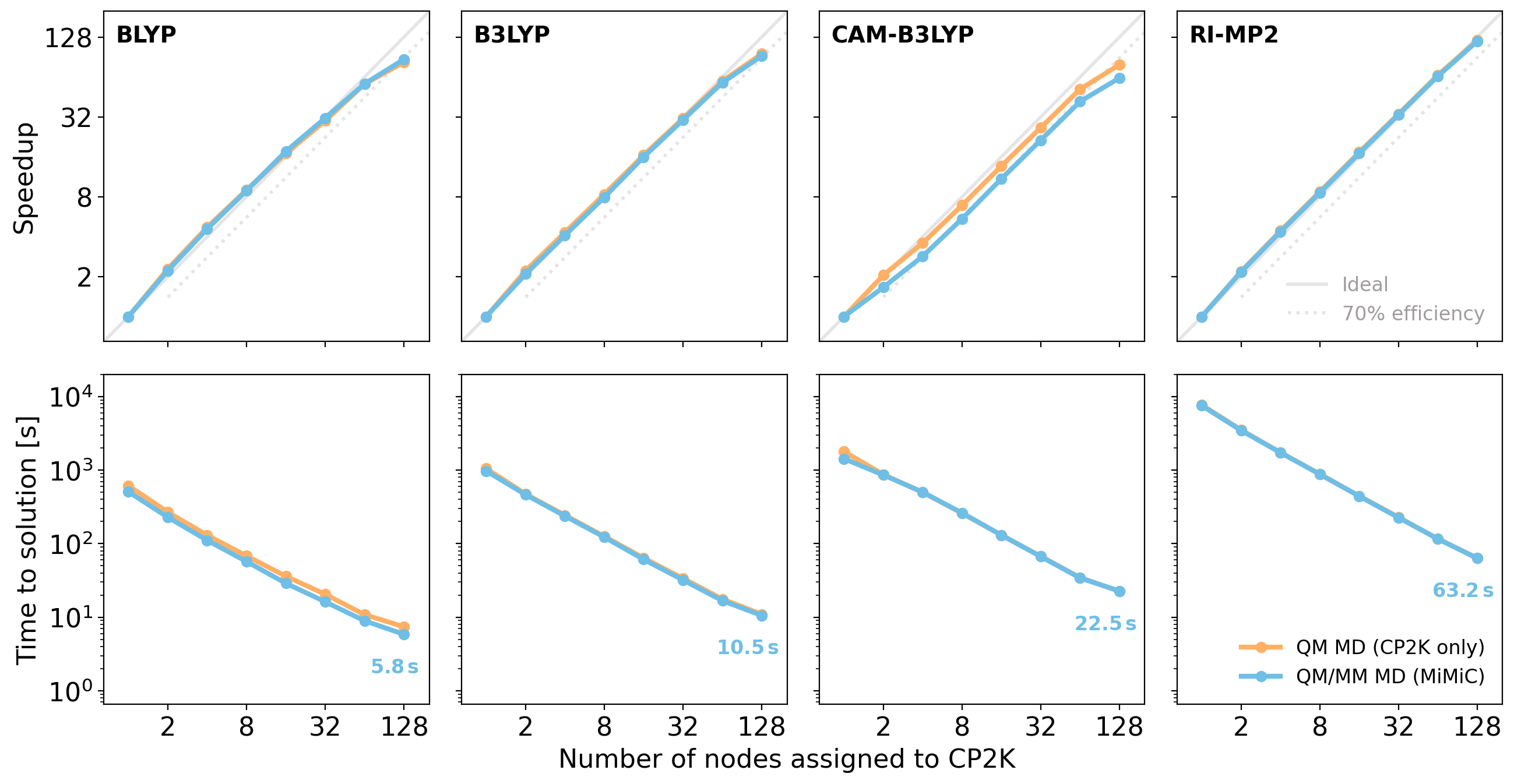}
    \caption{Scaling plots that show speedups and time to solution for different methods with an increasing amount of resources allocated to CP2K.
    Measured times correspond to the time used to solve the QM equations and to evaluate the QM forces in MiMiC-based QM/MM MD simulations of GFP(aq) (blue), i.e., only the parts of the calculations that are fully independent of MiMiC (omitting gathering/scattering of distributed data and transfers).
    These are compared against times from stand-alone CP2K ab initio MD runs of the isolated QM subsystem (orange).
    Speedups are calculated with respect to a single-node simulation using 2 MPI processes per node for CP2K.
    Thick and thin gray lines indicate ideal speedup and 70\% efficiency, respectively.
    Fastest times to solution values are shown near their respective points.}
    \label{fig:cp2k_scaling}
\end{figure}

The plots in Figure \ref{fig:cp2k_scaling} demonstrate the parallel scaling of different methods in CP2K expressed in terms of speedup and time to solution with respect to an increasing amount of computational resources allocated to the program.
From these, it can be seen that in all cases, CP2K performs nearly identically regardless of whether it was used in a QM/MM MD context or for a stand-alone ab initio MD simulation.
The scaling, expressed in terms of speedups, drops below \num{70}\% parallel efficiency only at 128 nodes for all methods except for MP2, which maintains high parallel efficiency and could potentially scale even further.
In the CAM-B3LYP speedup plot, the two lines differ, yet, after inspection of the respective time to solution, it becomes apparent that this happens due to slightly better single-node performance in a MiMiC-based simulation.
An interesting observation is that with the BLYP functional, the simulation using MiMiC appears to be more efficient, with the time to solution being consistently lower compared to the CP2K MD run.
This can be attributed to the fact that the system was equilibrated by performing QM/MM MD simulations with the BLYP functional, which resulted in fewer steps required to converge the wave function for the QM/MM MD runs compared to the pure QM MD runs using a system consisting of only the QM subsystem.
We also observed the same difference in performance with the other methods, although not as pronounced as in the case involving BLYP.

\begin{figure}
    \centering
    \includegraphics[width=0.5\linewidth]{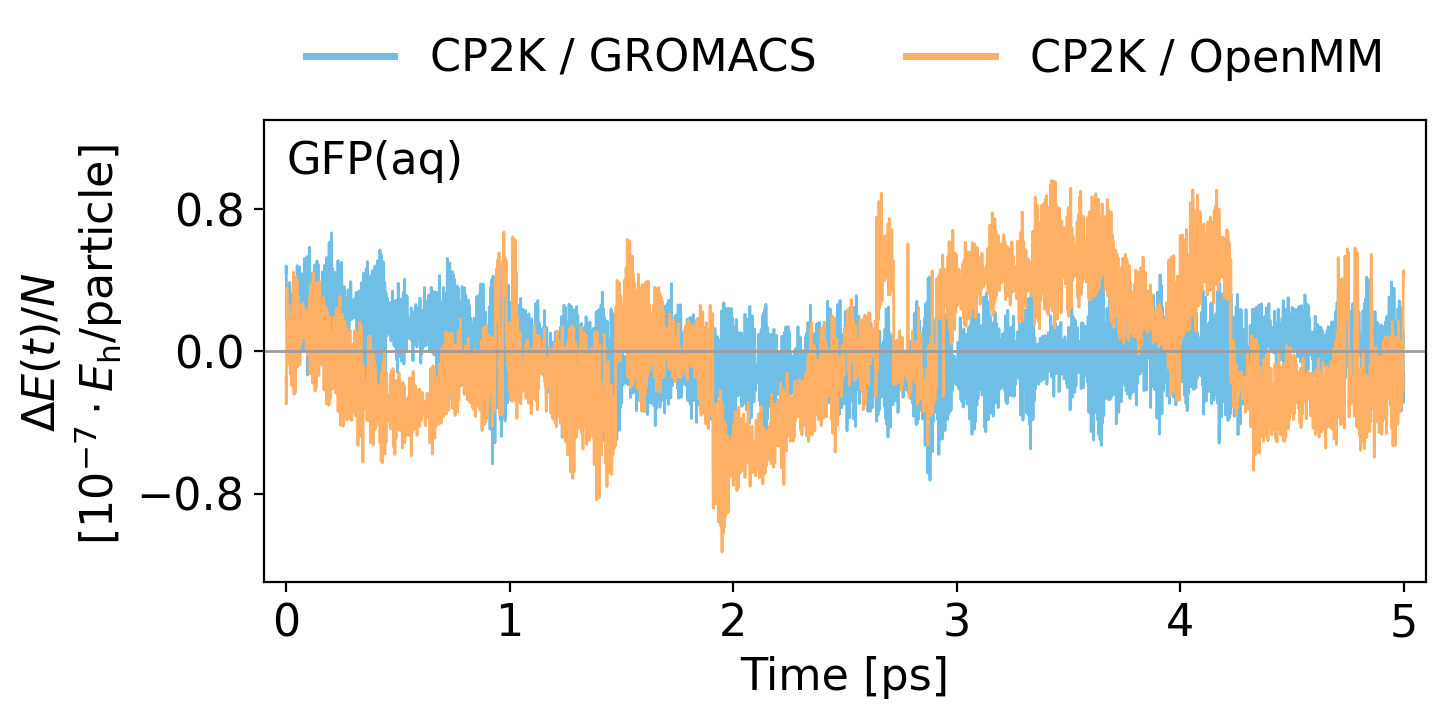}
    \caption{Fluctuations of the total energy per particle in a QM/MM MD simulation of GFP(aq) performed with MiMiC using CP2K for the QM subsystem, and either OpenMM (orange) or GROMACS (blue) for the MM subsystem.
    The change in energy per particle is given relative to the average energy ($\Delta E(t) = E(t) - \langle E \rangle$).}
    \label{fig:openmm}
\end{figure}

Finally, thanks to the efforts presented in this paper combined with the latest developments in MiMiC, we were able to fully realize its modular philosophy.
Namely, we recently introduced a new interface\cite{Levy2025} to OpenMM\cite{Eastman2023openmm}, opening the possibility to use it as an MM client program.
Now, with no additional implementation effort, we were able to replace one software component with another.
Instead of GROMACS, we could use OpenMM to perform simulations with CP2K just as efficiently, thus leading to a CP2K/OpenMM interface for QM/MM MD. 
We demonstrate the stability of this combination by performing a validation run on GFP(aq) with a small QM subsystem, the same as in Section~\ref{sec:valid}.
The plot in Figure~\ref{fig:openmm} shows the fluctuations of the total energy per particle with respect to the average energy.
Although it features larger fluctuations than the ones seen in Figure~\ref{fig:validation}, the total energy is well conserved with no drift. 

\section{Summary and Outlook}

% INTRO
The presented refactoring of MiMiC is a major step toward fully unlocking its originally intended flexibility, which effectively enables developers to easily introduce new external QM programs into the ecosystem.
To this end, we also implemented a new interface with CP2K and demonstrated that it is capable of performing efficient and stable QM/MM MD simulations.
Thanks to MiMiC's design, we could easily obtain a new simulation setup between CP2K and OpenMM without additional effort.
These developments also open the possibility to implement novel multiscale models such as QM/QM/MM and machine-learning-accelerated QM/MM.

% DATA TRANSFER
We performed a thorough timing analysis to demonstrate that communication between the programs does not create a bottleneck, despite the changes in MiMiC substantially increasing the amount of transferred data.
In fact, the framework remains highly efficient and the additional communication does not create a noticeable overhead that would result in decreased performance, at least within a reasonable range of parallel processes.
Furthermore, we show that the parallel performance of the client program is unhindered by MiMiC.
Still, the need to gather distributed quantities on a single process for their transfer between MiMiC and the external programs might become a bottleneck in the future for extremely distributed calculations involving thousands of MPI processes.
To address this, a more efficient communication scheme can be developed, in which different processes of client programs exchange data directly with different processes of the server program.

% RESOURCES AND SOLUTION
We demonstrated in the timing analysis that the majority of the computational time used by MiMiC is spent in calculating the short-range interactions, which is an embarrassingly parallel problem that can be drastically accelerated by employing GPUs.
In fact, we are currently working on offloading these interactions to GPUs, together with other parts of the code, to achieve considerable speed-up.
Nevertheless, we intend to optimize the code also for situations in which no GPU accelerators are available.
For these, we will reintroduce and optimize OpenMP threading, which was not taken into account in this work, as we wanted to focus on the effects of the distributed data transfer on the overall performance.

% FEATURES
Finally, we emphasize that the presented efforts revolve not only around CP2K but also make the framework more approachable, allowing developers to interface with new external programs.
This is particularly attractive because, compared to more tightly-coupled approaches for multiscale simulations, MiMiC offers much more flexibility while still maintaining high efficiency.
On top of this, thanks to the active development in MiMiC, many upcoming features will become available in the future.
These include the fit of D-RESP charges\cite{laio2002dresp, laio2004variational}, extended to multipoles, on-the-fly force field parametrization via force matching\cite{maurer2007automated, doemer2014generalized}, and an efficient fully polarizable QM/MM scheme \cite{Sonata2024}.

\section*{Author Information}
\subsection*{Notes}
The authors declare no competing financial interest.

\begin{acknowledgement}
The authors thank Vladislav Slama for assistance with setting up the CP2K calculations involving exact exchange.
This work has been supported by the Swiss National Science Foundation (Grants No. 200020-185092 and No. 200020-219440), and used computing time from the Swiss National Computing Centre CSCS.
J.M.H.O. gratefully acknowledges financial support from VILLUM FONDEN (Grant No. VIL29478).
\end{acknowledgement}

\section*{Associated Content}
The source code of the MiMiC framework is free and open-source and is hosted on GitLab\cite{mimic_project_gitlab}.
The data underlying this study and the source code of the MiMiC framework used at the time of data gathering are openly available in Zenodo at \url{https://doi.org/10.5281/zenodo.13940978}.

Supporting Information contains a detailed list of parameters used in simulations,
information about the used QM subsystems,
convergence analysis of the long-range electrostatic embedding scheme for the GFP(aq) system,
energy fluctuations of the full validation runs,
and timing data plotted in Figure~\ref{fig:data_transfer}.

{\footnotesize
\bibliography{references}
}

\onecolumn

\newpage
\section*{Supporting Information}
\input{SI.tex}

\end{document}

%% file: SI.tex
% Redefine the table and figure numbering to include the prefix "S"
\setcounter{figure}{0} 
\setcounter{table}{0} 
\renewcommand{\thetable}{S\arabic{table}}
\renewcommand{\thefigure}{S\arabic{figure}}

\newcommand*\mr[3]{\multirow{#1}{#2}{#3}}
\newcommand*\mc[3]{\multicolumn{#1}{#2}{#3}}
\renewcommand{\textfraction}{0.1}

%\section*{Detailed list of computational parameters}
\begin{table}
\centering
\caption{Total number of atoms in the MM subsystems, and sizes and shapes of the used simulation box sizes.}
\label{tab:gromacs_params}
\begin{tabular}{l|c|cl}
System    & Num. atoms   & Box type     & Dimension(s)            \\ \hline
Ace(aq)   & \num{2944}   & cubic        & \SI{3.10}{\nano\meter}  \\
GFP(aq)   & \num{28264}  & orthorhombic & \SI{7.44}{\nano\meter}, \SI{6.53}{\nano\meter}, \SI{6.12}{\nano\meter} \\
BuOH(ace) & \num{139495} & cubic        & \SI{11.94}{\nano\meter} \\ \hline
Small MM  & \num{2944}   & cubic        & \SI{3.10}{\nano\meter}  \\
Large MM  & \num{98314}  & cubic        & \SI{10.00}{\nano\meter}
\end{tabular}
\end{table}
\begin{table}
\centering
\caption{QM subsystem sizes and lists of CPMD and MiMiC parameters used in the validation simulations.
For CPMD, plane-wave basis cutoffs, wave-function extrapolation orders, and box sizes are listed.
For MiMiC, classical atoms were resorted into short- (sr) and long-range (lr) lists every 50 steps with a given sorting method, and the lr electrostatic interactions were controlled by two parameters: multipole expansion order of the QM subsystem charge density, and the cutoff radius for sr atoms.
Note that the sorting settings are system dependent and these were used also in simulations involving CP2K, i.e., also with the refactored development version of MiMiC.}
\label{tab:cpmd_mimic_params}
\begin{tabular}{l|c|ccc|ccc}
\mr{2}{*}{System} & \mr{2}{*}{Num. atoms} & \mc{3}{c|}{CPMD Parameters}
                  & \mc{3}{c}{MiMiC: sr/lr sorting} \\
       &            & Cutoff         & Extrap. & Box size       & Method    & Order   & Radius \\ \hline
Ace(aq)    & \num{10}   & \num{80}\,Ry   &  3      & \SI{40}{\bohr} & centroid  & \num{5} & \SI{30}{\bohr}\\
GFP(aq)    & \num{22}   & \num{115}\,Ry  &  3      & \SI{40}{\bohr} & atom-wise & \num{9} & \SI{40}{\bohr}\\
BuOH(ace)  & \num{15}   & \num{85}\,Ry   &  5      & \SI{40}{\bohr} & centroid  & \num{5} & \SI{32}{\bohr}
\end{tabular}
\end{table}
\begin{table}
\centering
\caption{Various sets of parameters used in the simulations with CP2K.
In all simulations, we used a 5-level multi-grid and further specified plane-wave cutoffs, relative cutoffs controlling the mapping of product Gaussians on different multi-grid levels, basis sets, and wave-function extrapolation orders.}
\label{tab:cp2k_params}
\begin{tabular}{l|cccc|ccc}
Settings & Cutoff      & Rel. cutoff & Basis & Extrap. & \mc{3}{c}{Box size} \\
         & [Ry]        & [Ry]        &       & order   & Ace & GFP & BuOH   \\ \hline
Loose    & \num{450}   & \num{50}    & DZVP  & 4 & \SI{10}{\angstrom} & \SI{13}{\angstrom} & \SI{10}{\angstrom}  \\
Tight    & \num{550}   & \num{60}    & TZVP  & 4 & \SI{20}{\bohr} & \SI{15}{\angstrom} & \SI{20}{\bohr} \\ \hline
Small QM & \num{400}   & \num{60}    & TZVP  & 4 & \SI{15}{\bohr} & $-$ & $-$  \\
Large QM & \num{1200}  & \num{60}    & TZVP  & 4 & \SI{30}{\bohr} & $-$ & $-$  
\end{tabular}
\end{table}

\begin{figure}
\centering
\includegraphics[width=0.8\linewidth]{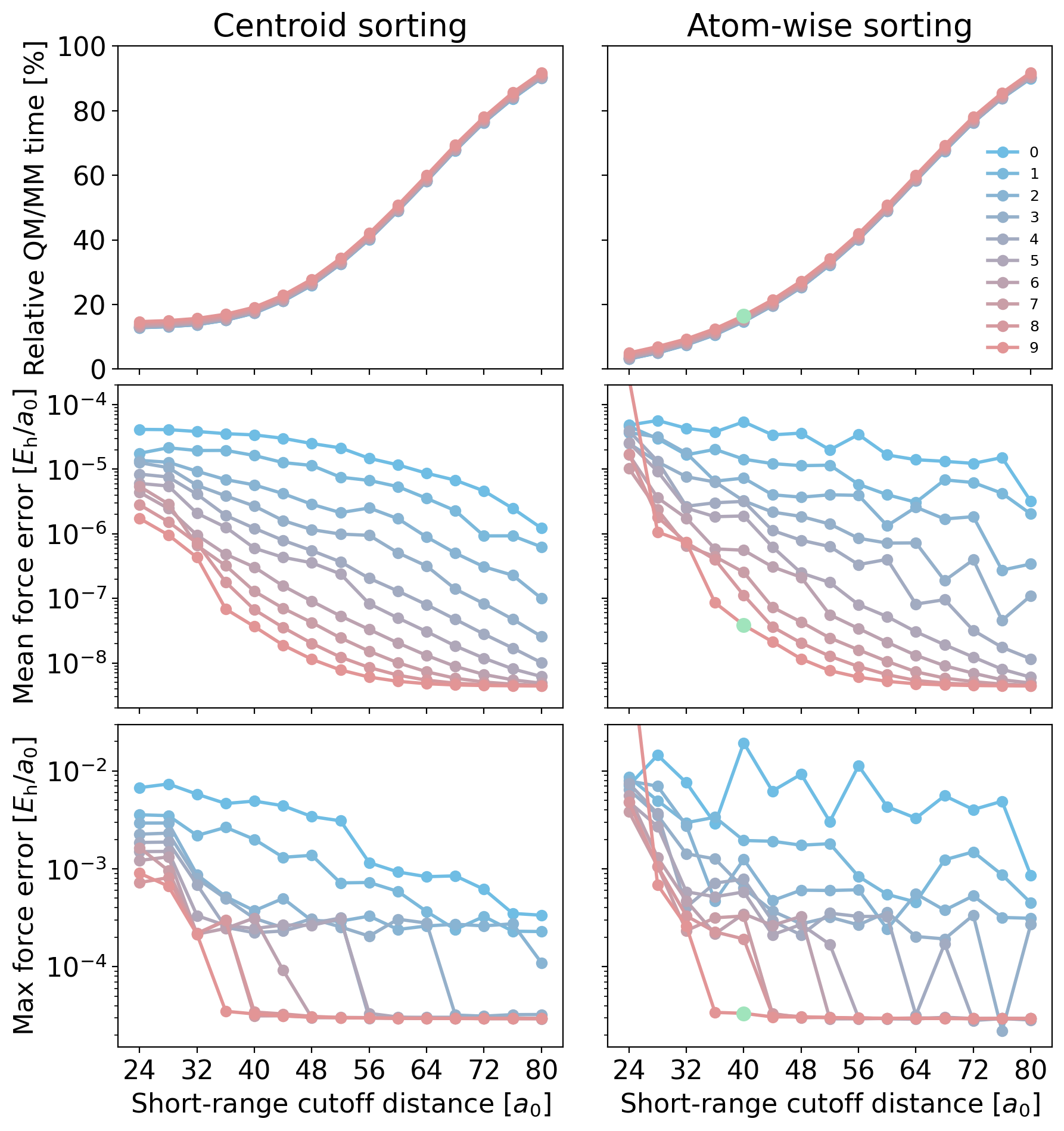}
\caption{Benchmarks of MiMiC long-range electrostatic interaction parameters for GFP(aq). Relative wall times, mean force errors, and maximum error in forces for calculating QM/MM terms with respect to the parameters controlling the long-range electrostatic interactions, i.e., the short-range cutoff distance and the order of the multipole expansion (indicated by different line colors).
The green marker denotes the final parameter values that were selected for the production runs.}
\label{fig:si_srlr_bechmarks}
\end{figure}

\begin{figure}
    \centering
    \includegraphics[width=\linewidth]{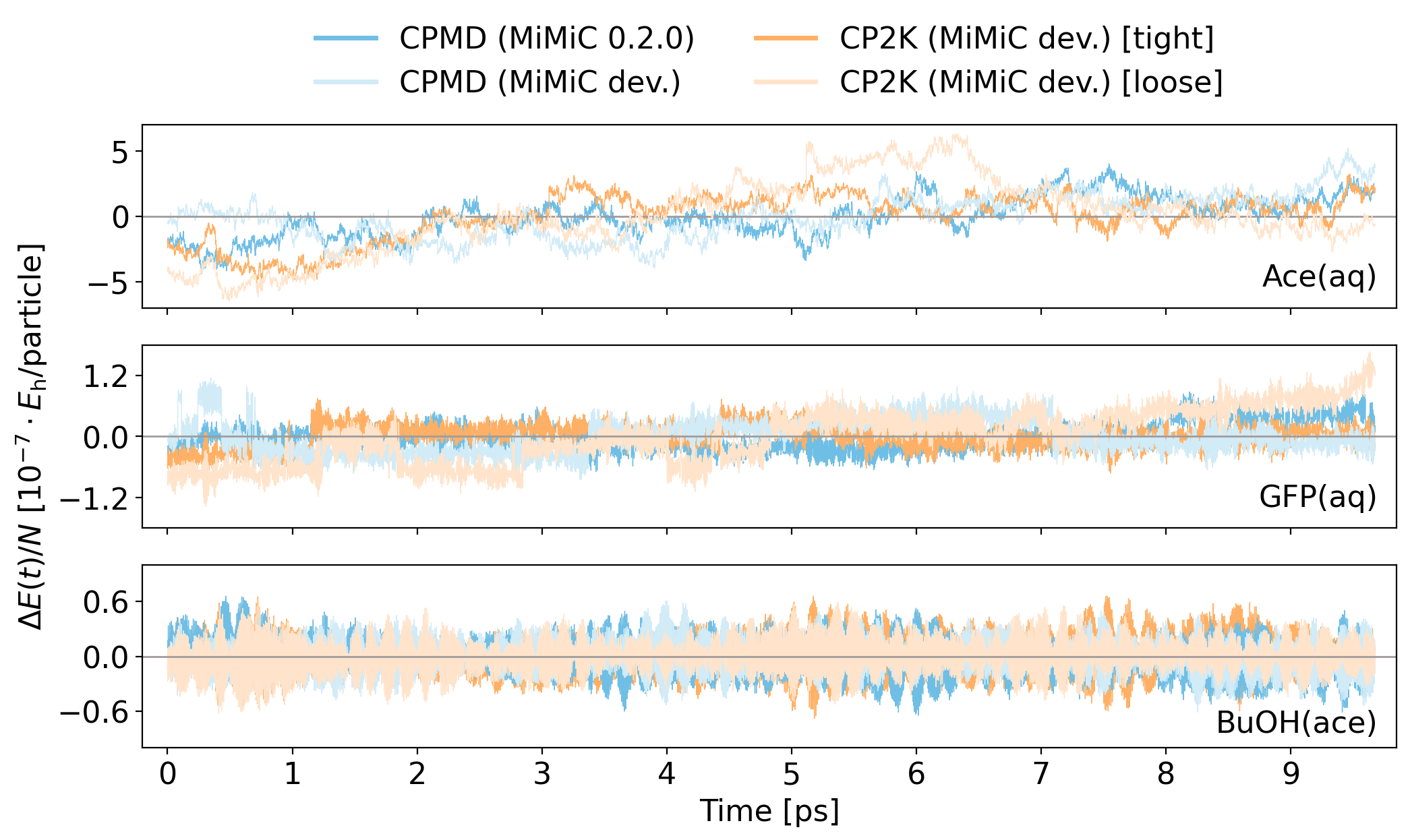}
    \caption{Fluctuations of the total energy per particle in the QM/MM MD simulations performed with MiMiC for Ace(aq), GFP(aq), and BuOH(ace).
    Full NVE trajectories from Figure~\ref{fig:validation}, i.e., including the initial $\sim$\SI{4.6}{\pico\second} that were considered to be an equilibration after switching off the thermostat.
    Additionally, we present simulations that used the looser settings for CP2K (see Table~\ref{tab:cp2k_params}) and those with CPMD, but with refactored MiMiC.
    }
    \label{fig:si_validation}
\end{figure}

\begin{figure}
    \centering
    \includegraphics[width=0.6\linewidth]{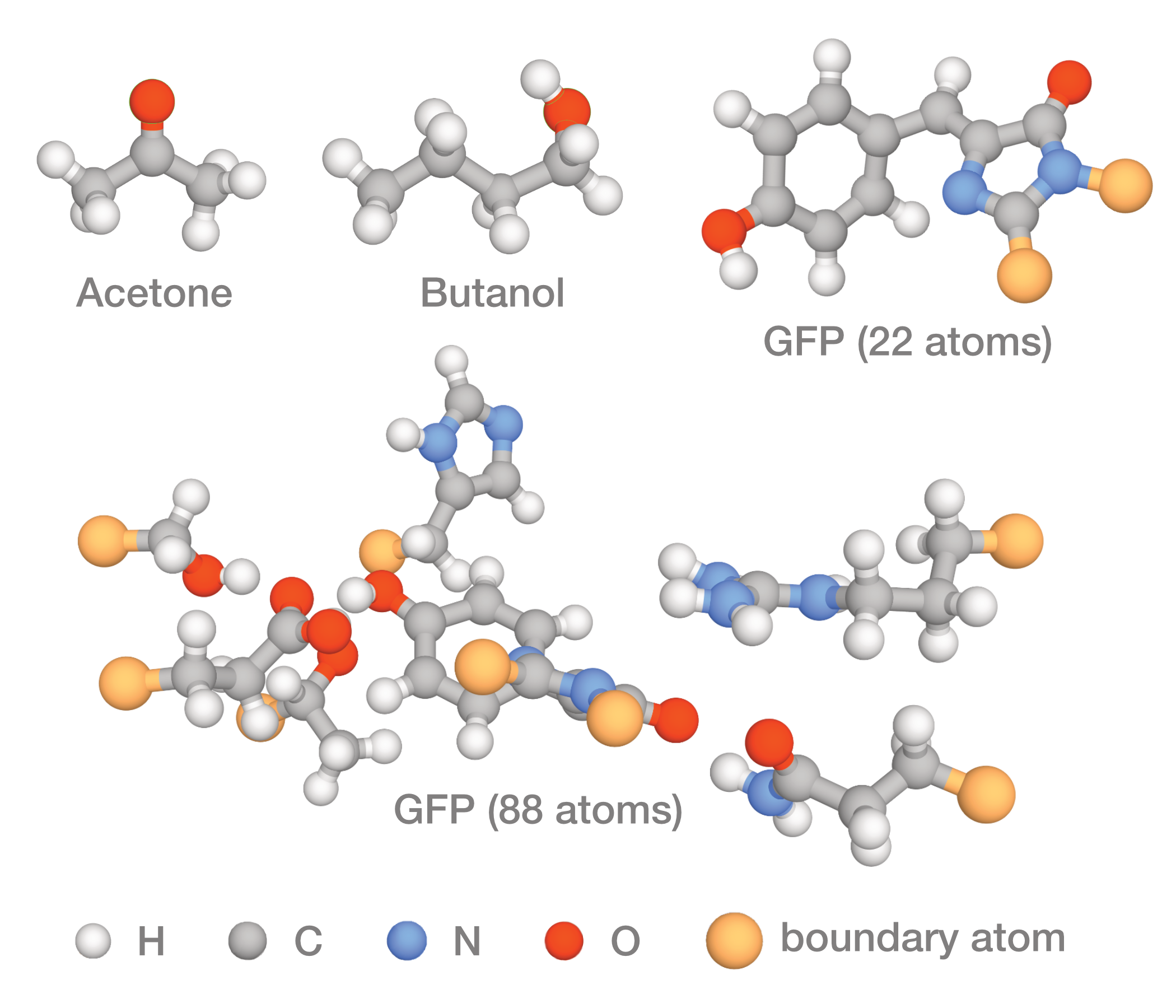}
    \caption{QM subsystems used in the simulations presented in the article.}
    \label{fig:si_systems}
\end{figure}

\begin{table}
\scalebox{0.62}{ \centerline{
\begin{tabular}{ccc|rcrc|rcrc|rcrc|rcrc|rc|r}
& \mc{2}{c|}{Num. procs} & \mc{4}{c|}{Long-range interactions} & \mc{4}{c|}{Short-range interactions} & \mc{4}{c|}{CP2K} & \mc{4}{c|}{Communication} & \mc{2}{c|}{Other} & \mc{1}{c}{Total} \\ 
& MiMiC & CP2K & \mc{2}{c}{tens.+mult.} & \mc{2}{c|}{pot.+forces} &
\mc{2}{c}{potential} & \mc{2}{c|}{forces}& \mc{2}{c}{SCF+forces} & \mc{2}{c|}{distributed} & \mc{2}{c}{distributed} & \mc{2}{c|}{transfer}& & & \\ \cline{2-22}
\mr{16}{*}{\rotatebox[origin=c]{90}{Small MM / Small QM}}
& 12 & 8 & 0.028 & (1.0) & 0.032 & (1.2) & 0.570 & (20.5) & 0.575 & (20.7) & 1.555 & (56.0) & 0.004 & (0.1) & 0.002 & (0.1) & 0.000 & (0.0) & 0.009 & (0.3) & 2.775 \\
& 12 & 16 & 0.028 & (1.3) & 0.032 & (1.4) & 0.570 & (25.6) & 0.575 & (25.8) & 1.008 & (45.3) & 0.003 & (0.1) & 0.002 & (0.1) & 0.000 & (0.0) & 0.009 & (0.4) & 2.228 \\
& 12 & 32 & 0.028 & (1.4) & 0.032 & (1.6) & 0.570 & (27.9) & 0.575 & (28.2) & 0.821 & (40.2) & 0.003 & (0.2) & 0.002 & (0.1) & 0.000 & (0.0) & 0.011 & (0.5) & 2.042 \\
& 12 & 64 & 0.028 & (1.4) & 0.032 & (1.6) & 0.570 & (27.6) & 0.575 & (27.9) & 0.842 & (40.8) & 0.004 & (0.2) & 0.002 & (0.1) & 0.000 & (0.0) & 0.009 & (0.4) & 2.062 \\
& 24 & 8 & 0.015 & (0.7) & 0.015 & (0.7) & 0.285 & (13.1) & 0.288 & (13.2) & 1.554 & (71.5) & 0.004 & (0.2) & 0.002 & (0.1) & 0.000 & (0.0) & 0.010 & (0.5) & 2.174 \\
& 24 & 16 & 0.015 & (0.9) & 0.015 & (0.9) & 0.286 & (17.5) & 0.288 & (17.7) & 1.013 & (62.1) & 0.003 & (0.2) & 0.002 & (0.1) & 0.000 & (0.0) & 0.009 & (0.5) & 1.631 \\
& 24 & 32 & 0.015 & (1.0) & 0.015 & (1.0) & 0.285 & (19.8) & 0.288 & (20.0) & 0.825 & (57.2) & 0.003 & (0.2) & 0.002 & (0.1) & 0.000 & (0.0) & 0.009 & (0.6) & 1.443 \\
& 24 & 64 & 0.016 & (1.1) & 0.015 & (1.0) & 0.285 & (19.7) & 0.288 & (19.9) & 0.824 & (57.0) & 0.004 & (0.3) & 0.002 & (0.1) & 0.000 & (0.0) & 0.012 & (0.8) & 1.446 \\
& 48 & 8 & 0.007 & (0.4) & 0.008 & (0.4) & 0.147 & (7.8) & 0.148 & (7.8) & 1.559 & (82.7) & 0.004 & (0.2) & 0.002 & (0.1) & 0.000 & (0.0) & 0.010 & (0.5) & 1.885 \\
& 48 & 16 & 0.007 & (0.5) & 0.008 & (0.6) & 0.147 & (10.5) & 0.148 & (10.6) & 1.074 & (76.6) & 0.004 & (0.3) & 0.004 & (0.3) & 0.000 & (0.0) & 0.010 & (0.7) & 1.402 \\
& 48 & 32 & 0.007 & (0.6) & 0.008 & (0.7) & 0.146 & (13.0) & 0.148 & (13.2) & 0.797 & (71.0) & 0.003 & (0.3) & 0.002 & (0.2) & 0.000 & (0.0) & 0.011 & (1.0) & 1.123 \\
& 48 & 64 & 0.007 & (0.6) & 0.008 & (0.7) & 0.147 & (13.4) & 0.148 & (13.5) & 0.773 & (70.3) & 0.003 & (0.3) & 0.002 & (0.2) & 0.000 & (0.0) & 0.010 & (1.0) & 1.099 \\
& 96 & 8 & 0.003 & (0.2) & 0.004 & (0.2) & 0.074 & (4.3) & 0.074 & (4.3) & 1.553 & (89.9) & 0.004 & (0.2) & 0.008 & (0.5) & 0.000 & (0.0) & 0.007 & (0.4) & 1.726 \\
& 96 & 16 & 0.003 & (0.3) & 0.004 & (0.3) & 0.074 & (6.2) & 0.074 & (6.2) & 1.018 & (85.5) & 0.003 & (0.3) & 0.006 & (0.5) & 0.000 & (0.0) & 0.008 & (0.7) & 1.191 \\
& 96 & 32 & 0.003 & (0.3) & 0.004 & (0.4) & 0.074 & (7.6) & 0.075 & (7.7) & 0.800 & (82.3) & 0.003 & (0.3) & 0.006 & (0.6) & 0.000 & (0.0) & 0.007 & (0.7) & 0.972 \\
& 96 & 64 & 0.003 & (0.3) & 0.004 & (0.4) & 0.074 & (7.6) & 0.074 & (7.6) & 0.799 & (82.0) & 0.004 & (0.4) & 0.008 & (0.8) & 0.000 & (0.0) & 0.008 & (0.9) & 0.974 \\
\cline{2-22}
\mr{16}{*}{\rotatebox[origin=c]{90}{Small MM / Large QM}}
& 48 & 36 & 0.453 & (0.9) & 0.425 & (0.9) & 10.119 & (20.4) & 9.841 & (19.8) & 28.098 & (56.6) & 0.218 & (0.4) & 0.232 & (0.5) & 0.003 & (0.0) & 0.238 & (0.5) & 49.626 \\
& 48 & 72 & 0.433 & (1.2) & 0.426 & (1.2) & 10.138 & (27.6) & 9.900 & (27.0) & 15.196 & (41.4) & 0.195 & (0.5) & 0.231 & (0.6) & 0.001 & (0.0) & 0.186 & (0.5) & 36.707 \\
& 48 & 144 & 0.508 & (1.7) & 0.427 & (1.4) & 10.064 & (34.1) & 9.880 & (33.4) & 8.018 & (27.1) & 0.192 & (0.7) & 0.231 & (0.8) & 0.001 & (0.0) & 0.217 & (0.7) & 29.538 \\
& 48 & 288 & 0.434 & (1.6) & 0.426 & (1.6) & 10.134 & (37.8) & 9.901 & (37.0) & 5.284 & (19.7) & 0.190 & (0.7) & 0.232 & (0.9) & 0.000 & (0.0) & 0.194 & (0.7) & 26.795 \\
& 96 & 36 & 0.225 & (0.6) & 0.215 & (0.6) & 5.099 & (13.0) & 4.985 & (12.8) & 27.944 & (71.5) & 0.219 & (0.6) & 0.224 & (0.6) & 0.003 & (0.0) & 0.173 & (0.4) & 39.087 \\
& 96 & 72 & 0.245 & (0.9) & 0.214 & (0.8) & 5.072 & (19.3) & 4.957 & (18.8) & 15.204 & (57.8) & 0.196 & (0.7) & 0.220 & (0.8) & 0.001 & (0.0) & 0.190 & (0.7) & 26.299 \\
& 96 & 144 & 0.225 & (1.2) & 0.215 & (1.1) & 5.095 & (26.5) & 4.987 & (26.0) & 8.070 & (42.0) & 0.195 & (1.0) & 0.224 & (1.2) & 0.001 & (0.0) & 0.185 & (1.0) & 19.196 \\
& 96 & 288 & 0.250 & (1.5) & 0.215 & (1.3) & 5.070 & (30.7) & 4.960 & (30.0) & 5.396 & (32.7) & 0.193 & (1.2) & 0.223 & (1.4) & 0.000 & (0.0) & 0.200 & (1.2) & 16.507 \\
& 192 & 36 & 0.129 & (0.4) & 0.107 & (0.3) & 2.467 & (7.3) & 2.480 & (7.3) & 28.077 & (82.9) & 0.220 & (0.6) & 0.215 & (0.6) & 0.003 & (0.0) & 0.190 & (0.6) & 33.888 \\
& 192 & 72 & 0.113 & (0.5) & 0.107 & (0.5) & 2.481 & (11.8) & 2.482 & (11.8) & 15.298 & (72.6) & 0.194 & (0.9) & 0.214 & (1.0) & 0.001 & (0.0) & 0.191 & (0.9) & 21.082 \\
& 192 & 144 & 0.121 & (0.9) & 0.107 & (0.8) & 2.476 & (17.9) & 2.495 & (18.0) & 8.055 & (58.2) & 0.192 & (1.4) & 0.212 & (1.5) & 0.001 & (0.0) & 0.181 & (1.3) & 13.839 \\
& 192 & 288 & 0.123 & (1.1) & 0.107 & (1.0) & 2.474 & (22.0) & 2.480 & (22.0) & 5.458 & (48.5) & 0.190 & (1.7) & 0.215 & (1.9) & 0.000 & (0.0) & 0.205 & (1.8) & 11.253 \\
& 384 & 36 & 0.065 & (0.2) & 0.055 & (0.2) & 1.200 & (3.9) & 1.211 & (3.9) & 27.927 & (89.8) & 0.220 & (0.7) & 0.216 & (0.7) & 0.003 & (0.0) & 0.187 & (0.6) & 31.084 \\
& 384 & 72 & 0.063 & (0.3) & 0.055 & (0.3) & 1.203 & (6.6) & 1.214 & (6.6) & 15.219 & (82.9) & 0.195 & (1.1) & 0.214 & (1.2) & 0.001 & (0.0) & 0.189 & (1.0) & 18.354 \\
& 384 & 144 & 0.069 & (0.6) & 0.055 & (0.5) & 1.199 & (10.7) & 1.216 & (10.9) & 8.043 & (71.9) & 0.191 & (1.7) & 0.214 & (1.9) & 0.001 & (0.0) & 0.193 & (1.7) & 11.181 \\
& 384 & 288 & 0.066 & (0.8) & 0.055 & (0.7) & 1.200 & (14.3) & 1.213 & (14.4) & 5.276 & (62.7) & 0.190 & (2.3) & 0.215 & (2.6) & 0.000 & (0.0) & 0.199 & (2.4) & 8.415 \\
\cline{2-22}
\mr{16}{*}{\rotatebox[origin=c]{90}{Large MM / Small QM}}
& 12 & 8 & 0.174 & (4.9) & 0.358 & (10.2) & 0.565 & (16.0) & 0.570 & (16.2) & 1.620 & (46.0) & 0.004 & (0.1) & 0.002 & (0.1) & 0.004 & (0.1) & 0.224 & (6.4) & 3.521 \\
& 12 & 16 & 0.174 & (6.0) & 0.358 & (12.3) & 0.564 & (19.4) & 0.570 & (19.6) & 1.002 & (34.5) & 0.003 & (0.1) & 0.002 & (0.1) & 0.004 & (0.1) & 0.225 & (7.8) & 2.903 \\
& 12 & 32 & 0.174 & (6.4) & 0.358 & (13.2) & 0.564 & (20.8) & 0.570 & (21.1) & 0.806 & (29.8) & 0.003 & (0.1) & 0.002 & (0.1) & 0.002 & (0.1) & 0.228 & (8.4) & 2.707 \\
& 12 & 64 & 0.174 & (6.4) & 0.358 & (13.1) & 0.564 & (20.7) & 0.570 & (20.9) & 0.821 & (30.2) & 0.004 & (0.1) & 0.002 & (0.1) & 0.004 & (0.1) & 0.227 & (8.3) & 2.724 \\
& 24 & 8 & 0.087 & (3.3) & 0.178 & (6.7) & 0.282 & (10.7) & 0.285 & (10.8) & 1.577 & (59.6) & 0.004 & (0.1) & 0.002 & (0.1) & 0.004 & (0.2) & 0.227 & (8.6) & 2.646 \\
& 24 & 16 & 0.087 & (4.1) & 0.178 & (8.4) & 0.282 & (13.3) & 0.285 & (13.4) & 1.051 & (49.5) & 0.003 & (0.2) & 0.002 & (0.1) & 0.002 & (0.1) & 0.233 & (11.0) & 2.123 \\
& 24 & 32 & 0.088 & (4.7) & 0.179 & (9.5) & 0.283 & (15.1) & 0.285 & (15.2) & 0.808 & (43.0) & 0.003 & (0.2) & 0.002 & (0.1) & 0.004 & (0.2) & 0.228 & (12.1) & 1.880 \\
& 24 & 64 & 0.087 & (4.6) & 0.179 & (9.4) & 0.282 & (14.8) & 0.285 & (15.0) & 0.826 & (43.5) & 0.004 & (0.2) & 0.002 & (0.1) & 0.002 & (0.1) & 0.233 & (12.3) & 1.900 \\
& 48 & 8 & 0.044 & (2.0) & 0.090 & (4.0) & 0.145 & (6.4) & 0.147 & (6.5) & 1.585 & (70.4) & 0.004 & (0.2) & 0.003 & (0.1) & 0.004 & (0.2) & 0.230 & (10.2) & 2.252 \\
& 48 & 16 & 0.044 & (2.6) & 0.090 & (5.4) & 0.145 & (8.6) & 0.147 & (8.8) & 1.013 & (60.3) & 0.003 & (0.2) & 0.003 & (0.2) & 0.002 & (0.1) & 0.231 & (13.8) & 1.678 \\
& 48 & 32 & 0.045 & (3.1) & 0.090 & (6.2) & 0.145 & (10.0) & 0.147 & (10.1) & 0.793 & (54.5) & 0.003 & (0.2) & 0.003 & (0.2) & 0.004 & (0.3) & 0.226 & (15.5) & 1.456 \\
& 48 & 64 & 0.044 & (3.1) & 0.090 & (6.3) & 0.145 & (10.1) & 0.147 & (10.2) & 0.774 & (53.8) & 0.003 & (0.2) & 0.002 & (0.1) & 0.002 & (0.1) & 0.231 & (16.1) & 1.438 \\
& 96 & 8 & 0.021 & (1.0) & 0.044 & (2.2) & 0.073 & (3.6) & 0.074 & (3.6) & 1.579 & (77.6) & 0.004 & (0.2) & 0.008 & (0.4) & 0.004 & (0.2) & 0.229 & (11.2) & 2.035 \\
& 96 & 16 & 0.021 & (1.4) & 0.044 & (3.0) & 0.073 & (4.9) & 0.074 & (5.0) & 1.019 & (69.0) & 0.003 & (0.2) & 0.007 & (0.5) & 0.004 & (0.3) & 0.231 & (15.6) & 1.476 \\
& 96 & 32 & 0.021 & (1.7) & 0.044 & (3.5) & 0.073 & (5.9) & 0.074 & (5.9) & 0.795 & (63.7) & 0.003 & (0.3) & 0.007 & (0.6) & 0.002 & (0.2) & 0.228 & (18.3) & 1.247 \\
& 96 & 64 & 0.021 & (1.7) & 0.044 & (3.5) & 0.073 & (5.8) & 0.074 & (5.9) & 0.792 & (63.4) & 0.004 & (0.3) & 0.008 & (0.6) & 0.004 & (0.3) & 0.231 & (18.4) & 1.250 \\
\cline{2-22}
\mr{16}{*}{\rotatebox[origin=c]{90}{Large MM / Large QM}}
& 48 & 36 & 0.476 & (0.9) & 0.508 & (1.0) & 9.835 & (19.5) & 9.590 & (19.0) & 29.247 & (57.9) & 0.217 & (0.4) & 0.233 & (0.5) & 0.007 & (0.0) & 0.444 & (0.9) & 50.557 \\
& 48 & 72 & 0.474 & (1.3) & 0.507 & (1.4) & 9.843 & (26.5) & 9.638 & (25.9) & 15.849 & (42.7) & 0.195 & (0.5) & 0.232 & (0.6) & 0.005 & (0.0) & 0.404 & (1.1) & 37.147 \\
& 48 & 144 & 0.510 & (1.7) & 0.507 & (1.7) & 9.806 & (33.1) & 9.615 & (32.4) & 8.368 & (28.2) & 0.191 & (0.6) & 0.232 & (0.8) & 0.005 & (0.0) & 0.432 & (1.5) & 29.666 \\
& 48 & 288 & 0.466 & (1.7) & 0.508 & (1.9) & 9.853 & (36.7) & 9.638 & (35.9) & 5.537 & (20.6) & 0.188 & (0.7) & 0.233 & (0.9) & 0.004 & (0.0) & 0.417 & (1.6) & 26.845 \\
& 96 & 36 & 0.237 & (0.6) & 0.256 & (0.6) & 4.957 & (12.3) & 4.852 & (12.0) & 29.175 & (72.3) & 0.217 & (0.5) & 0.225 & (0.6) & 0.007 & (0.0) & 0.403 & (1.0) & 40.328 \\
& 96 & 72 & 0.239 & (0.9) & 0.255 & (0.9) & 4.958 & (18.3) & 4.833 & (17.9) & 15.916 & (58.9) & 0.197 & (0.7) & 0.223 & (0.8) & 0.005 & (0.0) & 0.409 & (1.5) & 27.035 \\
& 96 & 144 & 0.237 & (1.2) & 0.256 & (1.3) & 4.957 & (25.3) & 4.850 & (24.8) & 8.438 & (43.1) & 0.194 & (1.0) & 0.227 & (1.2) & 0.005 & (0.0) & 0.410 & (2.1) & 19.574 \\
& 96 & 288 & 0.238 & (1.4) & 0.256 & (1.5) & 4.958 & (29.6) & 4.834 & (28.8) & 5.641 & (33.6) & 0.194 & (1.2) & 0.226 & (1.3) & 0.004 & (0.0) & 0.422 & (2.5) & 16.774 \\
& 192 & 36 & 0.138 & (0.4) & 0.128 & (0.4) & 2.402 & (6.8) & 2.419 & (6.9) & 29.310 & (83.1) & 0.216 & (0.6) & 0.223 & (0.6) & 0.011 & (0.0) & 0.407 & (1.2) & 35.254 \\
& 192 & 72 & 0.136 & (0.6) & 0.128 & (0.6) & 2.407 & (11.0) & 2.422 & (11.1) & 15.932 & (72.9) & 0.195 & (0.9) & 0.222 & (1.0) & 0.005 & (0.0) & 0.410 & (1.9) & 21.858 \\
& 192 & 144 & 0.128 & (0.9) & 0.128 & (0.9) & 2.415 & (16.9) & 2.433 & (17.0) & 8.396 & (58.6) & 0.192 & (1.3) & 0.220 & (1.5) & 0.005 & (0.0) & 0.406 & (2.8) & 14.323 \\
& 192 & 288 & 0.129 & (1.1) & 0.128 & (1.1) & 2.412 & (20.8) & 2.420 & (20.8) & 5.682 & (48.9) & 0.189 & (1.6) & 0.223 & (1.9) & 0.004 & (0.0) & 0.428 & (3.7) & 11.615 \\
& 384 & 36 & 0.064 & (0.2) & 0.065 & (0.2) & 1.177 & (3.6) & 1.188 & (3.7) & 29.164 & (89.7) & 0.216 & (0.7) & 0.233 & (0.7) & 0.007 & (0.0) & 0.394 & (1.2) & 32.508 \\
& 384 & 72 & 0.071 & (0.4) & 0.065 & (0.3) & 1.169 & (6.1) & 1.179 & (6.1) & 15.924 & (82.7) & 0.195 & (1.0) & 0.231 & (1.2) & 0.005 & (0.0) & 0.412 & (2.1) & 19.252 \\
& 384 & 144 & 0.068 & (0.6) & 0.065 & (0.6) & 1.176 & (10.0) & 1.179 & (10.0) & 8.438 & (71.7) & 0.194 & (1.6) & 0.231 & (2.0) & 0.005 & (0.0) & 0.417 & (3.5) & 11.772 \\
& 384 & 288 & 0.067 & (0.8) & 0.066 & (0.7) & 1.171 & (13.3) & 1.183 & (13.4) & 5.502 & (62.4) & 0.190 & (2.2) & 0.231 & (2.6) & 0.004 & (0.0) & 0.408 & (4.6) & 8.823 \\
\cline{2-22}
\end{tabular}
} }
\caption{Timing data used to create the donut charts in Figure~\ref{fig:data_transfer}. All times shown in seconds, with the percentage of total time per MD step in parentheses.}
\label{tab:pie_chart_data}
\end{table}